\documentclass[11pt]{article}

\usepackage[utf8]{inputenc} 
\usepackage[T1]{fontenc} 
\usepackage{fouriernc} 
\usepackage[main=english,turkish]{babel}
\usepackage{csquotes}

\usepackage[mathscr]{eucal}

\providecommand{\keywords}[1]
{
  \small	
  \textbf{\textit{Keywords---}} #1
}

\usepackage{url}            
\usepackage{booktabs}       
\usepackage{amsfonts}       
\usepackage{nicefrac}       
\usepackage{microtype}      
\usepackage{lipsum}

\usepackage{geometry}        
\usepackage{natbib}
\usepackage{graphicx}
\usepackage{amssymb}
\usepackage{epstopdf}
\usepackage{xcolor}
\usepackage{comment}
\usepackage{enumitem}
\usepackage{epigraph}

\usepackage{mathtools}

\usepackage[printonlyused]{acronym}

\acrodef{aa}[a.a.]{almost all}
\acrodef{adpra}[ADPRA]{Advanced Dynamic Probabilistic Risk Assessment}
\acrodef{aec}[AEC]{Atomic Energy Commission}
\acrodef{as}[a.s.]{almost surely}
\acrodef{ba}[asm]{by assumption}
\acrodef{bwr}[BWR]{Boiling Water Reactor}
\acrodef{cba}[CBA]{Cost--Benefit Analysis}
\acrodef{cdf}[CDF]{Core Damage Frequency}
\acrodef{dpra}[DPRA]{Dynamic Probabilistic Risk Assessment}
\acrodef{loa}[LOA]{Lack of Anticipation}
\acrodef{ler}[LER]{Licensee Event Report}
\acrodef{fmea}[FMEA]{Failure Modes and Effects Analyses}
\acrodef{lopa}[LOPA]{Layers of Protection Analysis}
\acrodef{bdb}[BDB]{Beyond Design Basis}
\acrodef{bwr}[BWR]{Boiling Water Reactor}
\acrodef{bsee}[BSEE]{U.S. Bureau of Safety and Environmental Enforcement}
\acrodef{epri}[EPRI]{Electric Power Research Institute}
\acrodef{epa}[EPA]{Environmental Protection Agency}
\acrodef{cap}[CAP]{Condition Adverse to Quality Program}
\acrodef{cba}[CBA]{Cost-Benefit Analysis}
\acrodefplural{cba}[CBAs]{Cost-Benefit Analyses}
\acrodef{cdf}[CDF]{Core Damage Frequency}
\acrodef{chrs}[CHRS]{Containment Heat Removal System}
\acrodef{crmp}[CRMP]{Comprehensive Risk Management Process}
\acrodefplural{crmp}[CRMPs]{Comprehensive Risk Management Processes}

\acrodef{doe}[DOE]{Department of Energy}
\acrodef{db}[DB]{Design-basis}
\acrodef{dba}[DBA]{Design Basis Accident}
\acrodef{did}[DID]{Defense-in-Depth}
\acrodef{eccs}[ECCS]{Emergency Core Cooling System}
\acrodef{emrald}[EMRALD]{Event Modeling Risk Assessment using Linked Diagrams}
\acrodef{esd}[ESD]{Event Sequence Diagram}
\acrodef{flc}[FC]{Fails Closed}
\acrodef{fda}[FDA]{Food and Drug Administration}
\acrodef{fmea}[FMEA]{Failure Mode and Effects Analysis}
\acrodefplural{fmea}[FMEA]{Failure Modes and Effects Analyses}
\acrodef{fsar}[FSAR]{Final Safety Analysis Report}
\acrodef{fs}[FoS]{Factor of Safety}
\acrodefplural{fs}[FsoS]{Factors of Safety}
\acrodef{ft}[FT]{Fault Tree}
\acrodef{fts}[FTS]{Fail To Start}
\acrodef{ftr}[FTR]{Fail To Run}
\acrodef{gra}[GRA]{Generation Risk Assessment}
\acrodef{iid}[i.i.d]{independent and identically distributed}
\acrodef{inl}[INL]{Idaho National Laboratory}
\acrodef{lar}[LAR]{License Amendment Requests}
\acrodef{lb}[LB]{Licensing Basis}
\acrodef{ler}[LER]{Licensee Event Report}
\acrodef{lerf}[LERF]{Large Early Release Frequency}
\acrodef{loa}[LOA]{Lack of Anticipation}
\acrodef{loca}[LOCA]{Loss of Coolant Accident}
\acrodef{lwr}[LWR]{Light Water Reactor}
\acrodef{mss}[MSS]{Main Steam System}
\acrodef{mslb}[MSLB]{Main Steam Line Break}
\acrodef{nei}[NEI]{Nuclear Energy Institute}
\acrodef{neima}[NEIMA]{Nuclear Energy Innovation and Modernization Act} 
\acrodef{npv}[NPV]{Net Present Value}
\acrodef{npp}[NPP]{Nuclear Power Plant}
\acrodef{nrc}[NRC]{Nuclear Regulatory Commission}
\acrodef{nasa}[NASA]{National Aeronautics and pace Administration}
\acrodef{oqap}[OQAP]{Operations Quality Assurance Program}
\acrodef{omb}[OMB]{Office of Management and Budget}
\acrodef{oira}[OIRA]{Office of Information and Regulatory Affairs}
\acrodef{om}[O\&M]{Operations and Maintenance}
\acrodef{ora}[ORA]{Organizational Risk Assessment}
\acrodef{pasta}[PASTA]{Poisson Arrivals See Time Averages}
\acrodef{pga}[PGA]{Peak Ground Acceleration}
\acrodef{pra}[PRA]{Probabilistic Risk Assessment}
\acrodef{psa}[PSA]{Probabilistic Safety Analysis}
\acrodef{pwr}[PWR]{Pressurized Water Reactor}
\acrodef{qra}[QRA]{Quantitative Risk Assessment}
\acrodef{psa}[PSA]{Probabilistic Safety Analysis}
\acrodef{uq}[UQ]{Uncertainty Quantification}
\acrodef{pq}[PQ]{Probability Quantification}
\acrodef{qra}[QRA]{Quantitative Risk Analysis}
\acrodef{rcb}[RCB]{Reactor Containment Building}
\acrodef{rcd}[RCD]{Reactor Core Damage}
\acrodef{rcs}[RCS]{Reactor Coolant System}
\acrodef{rr}[RR]{Radiation Release}
\acrodef{ssc}[SSC]{Systems, Structures, and Components}
\acrodef{stm}[STM]{State Transition Matrix}
\acrodefplural{stm}[STMs]{State Transition Matrices}
\acrodef{stp}[STP]{South Texas Project}
\acrodef{ufsar}[UFSAR]{Updated Final Safety Analysis Report}
\acrodef{wp1}[w.p.1]{with probability 1}

\usepackage{cleveref}

\usepackage[printonlyused]{acronym}

\newtheorem{remark}{Remark}

\crefname{theorem}{theorem}{theorems}
\Crefname{Theorem}{Theorem}{Theorems}
\crefname{proposition}{proposition}{propositions}
\Crefname{Property}{Property}{Properties}
\crefname{corollary}{corollary}{corollaries}
\Crefname{Corollary}{Corollary}{Corollaries}
\crefname{definition}{definition}{definitions}
\Crefname{Definition}{Definition}{Definitions}
\crefname{lemma}{lemma}{lemmas}
\Crefname{Lemma}{Lemma}{Lemmas}
\crefname{exercise}{exercise}{exercises}
\Crefname{Exercise}{Exercise}{Exercises}
\crefname{remark}{remark}{remarks}
\Crefname{Remark}{Remark}{Remarks}
\crefname{example}{example}{example}
\Crefname{Example}{Example}{Examples}

\usepackage{setspace}

\title{A Note on Probability Quantification for Protective System Efficacy Analysis: Stochastic Dynamics, Information Flow, and Initiating Event Arrival Times}

\author{
 Martin Wortman, Ernest Kee, and  Pranav Kannan\\
  HazTechRisk.Org\\
  }

\begin{document}
\maketitle

\begin{abstract}
	\ac{pq} predictions of the efficacy of safety--critical protective systems is challenging. Yet, the popularity of \ac{pq} methodologies (\emph{e.g.,} \ac{pra}, \ac{qra} and \ac{psa}) is growing and can now be found written into regulatory rules. \ac{pq} in predictive modeling is attractive because of its grounding in probability theory.\footnote{We hold that probability theory is the only viable theory of uncertainty.} But, certain important safety related events are not probability--measurable which is problematic for risk--analytic methodologies that rely on \ac{pq} computations.
Herein, we identify why the dynamics of available information play an essential role in governing the fidelity of \ac{pq}, and why \ac{pq} in the analysis of safety--critical protective systems is limited by the un-measurability of certain critical events. We provide an historical example that provides a practical context for our observations.  Finally we discuss the implications of measurability for regulatory decision--making governed by recent nuclear industry legislation advocating increased use of risk informed, performance--based regulation for advanced reactor licensing.
\end{abstract}


\pagebreak
	\epigraph{\emph{But there are also unknown unknowns --- the ones we don't know we don't know.}}{— \emph{Donald Rumsfeld}}

\section{Introduction}\label{introduction}

	Protective systems are designed to mitigate the consequences of failure in the hazardous technologies that they overlay.
	Typically, protections for technologies such as nuclear power plants, chemical processing facilities, and hazardous material transportation and storage that would ensure public safety are enacted in law and enforced by regulators.  
	Regulators are well aware that any protection can break down unexpectedly due to previously undiscovered failure modes.
	Failure modes that are first discovered under circumstances that, if not immediately addressed, might exceed to catastrophe are especially troubling. 
	Extensive records are kept on such ``near miss'' incidents, for example the \ac{nrc} \ac{ler} database which documents many such failures found in many different designs.
	Following near miss incidents, root cause analysis is used to isolate the immediate cause for the breakdown which can determine if it arose from something that can be assigned to physics, improper operation, or improper maintenance. Once isolated, the root cause is addressed by corrective action intended to prevent future occurrence.
	
	Knowing that unanticipated failure modes will be experienced, safety margin and defense in depth are used as primary protections against dangers that may be present.
	Protection designers use logic structures, such as \ac{fmea}, HazOp, fault trees, and event trees, to help study scenarios that have little, or no redundancy and may require additional measures of safety margin or defense in depth.
	If a single point of failure cannot be avoided, regulators are reluctant to accept a protection design including them unless the device where it is located has been tested thoroughly and includes substantial safety margin.
	Regular inspections help ensure the design retains safety margin over the device service life.
	
	%
	%
	We use the terminology unanticipated failure modes to indicate failure modes can only be found at a future time.
	The potential for their appearance in service is well understood by both protection designers and regulators.
	Prior to release of potentially hazardous processes or potentially hazardous products,
	good regulatory practice dictates a cautious approach with a gradual ramp up to full production.
	%
	
	%
	%
	Effective maintenance organizations address unexpected failures in critical equipment by aggressively pursuing
	root cause determination followed up with prompt corrective action.
	Although \ac{pq} works with probabilities of functional failures, the functional failures are
	collections of one or more failure modes including any unexpected ones.
	The importance of unanticipated failure modes is that they cannot be included in quantitative assessments such
	as \ac{pra}.
	
	The 2019 \ac{neima} asks for, in part, ``risk-informed'' regulation by the \ac{nrc}.
	While the \ac{nrc} currently uses risk-informed decision-making to guide certain policy decisions as well
	as prioritizing inspection and enforcement, they define risk-informed to include prudent engineering
	practices that include defense in depth and safety margin.
	Reliance on such practices are captured in the \ac{nrc} definition of risk-informed decision-making \citep{nrc2013}:
	\begin{quote}\label{neima_language}
	    ``A `risk-informed' approach to regulatory decision-making represents a philosophy whereby risk insights are 
	    considered together with other factors to establish requirements that better focus licensee and regulatory attention on design and operational issues commensurate with their importance to health and safety. A ‘risk-informed’ approach enhances the traditional approach by: (a) allowing explicit consideration of a broader set of potential challenges to safety, (b) providing a logical means for prioritizing these challenges based on risk significance, operating experience, and/or engineering judgment, (c) facilitating consideration of a broader set of resources to defend against these challenges, (d) explicitly identifying and quantifying sources of uncertainty in the analysis, and (e) leading to better decision-making by providing a means to test the sensitivity of the results to key assumptions. Where appropriate, a risk-informed regulatory approach can also be used to reduce unnecessary conservatism in deterministic approaches, or can be used to identify areas with insufficient conservatism and provide the bases for additional requirements or regulatory actions.''
	\end{quote}
	The un--measurability of the arrival times of future unexpected failure modes reinforces the need for defense in depth and safety margin to be included to help ensure the unbounded level of risk not included in \ac{pq} is taken into account in protective system regulation.\footnote{For example, Title 10 of the Code of Federal Regulations Part 50, Appendix A ``General Design Criteria'' requires such prudent engineering practices.}

	Some of the authoritative studies and texts we have found that consider \ac{pq} approaches to protection efficacy analysis include \citep{7218239820110601,apostolakis1981pitfalls,edsjsr.4027125620090101,rasmussen1975,kaplan1981}.  None of them addresses unmeasurable events central to efficacy analysis.
	
	%
	%
	We now briefly review a relevant, non-consequential, example of an unexpected failure mode from 
	an operating nuclear power plant and then provide the analytical formalism clarifying why catastrophes 
	arising from unanticipated failure modes defy \ac{pq}.
	We conclude with a brief discussion asserting that over-reliance
	on \ac{pq} should be avoided in risk analyses of safety-critical protections.

\section{An example}
	On December 18, 1995, the \ac{stp} Unit~1 commercial nuclear reactor experienced an unexpected reactor trip due to a main and auxiliary 
	transformer lockout while operating at 100\% power as described in \cite{ler95-013-01}.
	The transformer lockouts triggered a series of protective system actuations:
	\begin{enumerate}
		\item The generator output breaker tripped open,
		\item The main turbine tripped, causing a reactor trip, and
		\item All four 13.8 kV auxiliary bus breakers opened resulting in a loss of offsite power to the A Train Engineered Safeguards Features Bus.
	\end{enumerate}
	The plant operators responded by entering the emergency operating procedure for reactor/turbine trip which required verification of subcritical reactor and control
	rods fully inserted.
	However, three control rods, designated F10, C9, and N7, representing 3 out of a total of 57 control rods required to fully insert, indicated they 
	had instead stopped their travel about 6 steps prior to full insertion.\footnote{One step corresponds to an increment of less than 1/2 of 
	an inch of vertical motion.}
	All the control rods that had stopped prior to complete insertion were operating in a new fuel design.
	
	Proper operation of the reactor control rods, including full insertion following reactor trip, is verified prior to full power operation after the reactor core is replaced during a refueling outage.
	In protective operation, the control rods are designed to fall to the bottom of the core under the force of gravity in two basic stages, a free fall stage followed by a braking stage.
	They are tested to verify they fall within a prescribed length of time, from time of release to the time they start braking.
	This test was performed successfully as normally done for the new core installed after the refueling outage prior to the event on December 18.\footnote{``Refueling
	outage'' refers to a regularly scheduled period of time when approximately 1/3 of the reactor core is replaced with new fuel.}

	\subsection{Root cause}
		Other similar events started occurring at other reactor plants after the \ac{stp} October~18 event 
		\citep{bul96-01}.
		Root cause was investigated at the \ac{stp} over the next few days and it was concluded that the fuel assemblies, which
		had been exposed to radiation in at least one fuel cycle were becoming susceptible to buckling as partially described by
		\citet{24wrsim}.
		The root cause was found by developing a physical process theory model that was fit to data.
		Investigators developed new understanding of certain aspects of the control rod fluid shear and a fuel assembly annealing process similar to
		a spring and viscous damper with hysteresis in series, to explain the physics
		\citet{kee2005chemically}, and \citet{kb1997}.
		Various unrelated control rod insertion events have been observed, root cause isolated, and the knowledge base for these 
		kinds of events continues to grow larger for example, \citep{ler02-12-82,ien86-68}.
	
	\subsection{Corrective and compensatory actions}	
		Because the root cause investigation revealed that the control rods would insert to a point where the incremental reactivity control 
		was very small, the consequence analysis showed that, due to safety margin included in their design, the control rods would continue 
		to meet their reactivity control design requirements with continued margin to safety as prescribed in regulation.
		To validate the root cause and consequence analysis, the \ac{stp} proposed a compensatory testing plan to the \ac{nrc} which was accepted and implemented.
		To fully address the root cause and consequence analysis, the \ac{stp} suggested a modification to the fuel assembly designs being purchased in the future 
		that was implemented by the vendor.
				
	\subsection{Observations}
		As described by \citet{ler95-013-01}, the reactor trip revealed the existence of a new failure mode of 
		fuel assemblies that was itself triggered by a previously unknown failure mode due to a pinched pilot wire induced by maintenance activity.
		The timing of the pilot wire failure mode revealed the fuel assembly failure mode at an unexpected time during full power operation.
		Annealing of in-vessel components is a well-known process and aspects were always included in, for example, fuel assembly and \ac{bwr} channel box designs.
		However the process of annealing on strain rate as load approaches Euler buckling that was analytically described at the Ringhals nuclear power plant was unknown to be a problem in \ac{pwr} fuel assemblies.
		The fluid shear process in the braking section of the fuel assembly had never been investigated, at least in the academic literature, until
		\ac{stp} made empirical observations and \citeauthor{kee2005chemically} derived the necessary equations now published in \citep[][Section 4.2.1]{kee2005chemically}.
		
		It can be said that the safety margin included in the as-designed reactivity requirements of the control rods prevented a return to critical with the control rods not fully
		inserted.
		This observation helps support the need for such prescriptive regulations as those included in Title 10 CFR Part 50 that require safety margins and
		defense in depth for safety-critical functions.
		Root cause analysis wisely acknowledges the possible existence of unanticipated failure modes with plans to learn from their discovery.
	
\section{Protection Availability}\label{availability}

Our observations about \ac{pq} for protections are made with respect to a filtered probability space $(\Omega, \mathscr{F}, \{\mathscr{F}_t\}_{t \ge 0}, P)$, where $\Omega$ is the set of possible outcomes for a protective system over its lifecycle, $\mathscr{F}$ is a $\sigma$-algebra on $\Omega$ capturing all events of predictive modeling interest, and $P$ is a probability of the measurable space $(\Omega, \mathscr{F})$. The filtration $ \{\mathscr{F}_t\}_{t \ge 0}$ is the analytical construct used to capture the flow of engineering information over time. $ \{\mathscr{F}_t\}_{t \ge 0}$ contains all $P$-measurable collections of outcomes in $\Omega$ known to modelers by time $t \ge 0$. This filtration has the standard properties where for all $s,t \ge 0$,
	\begin{enumerate}
		\item $\mathscr{F}_0$ contains all $P$--null sets (completeness)\label{completeness},
		\item $\lim_{s \downarrow 0} \mathscr{F}_{t + s} \triangleq \bigcap_{s > t} \mathscr{F}_{s} = \mathscr{F}_{t}$ (right--continuity)\label{right-continuity},
		\item $\mathscr{F}_t \subseteq \mathscr{F}_{t+s}$ (monotonicity)\label{monotonicity},
		\item $\lim_{t \rightarrow \infty} \mathscr{F}_t \triangleq \bigvee_{t \ge 0} \mathscr{F}_t = \mathscr{F}$ (convergence)\label{convergence}.
	\end{enumerate}
From the perspective of protection system predictive modeling, these properties are normative and intuitively understandable. Property \ref{right-continuity} indicates that engineering information is not necessarily pre-visible (\emph{e.g.,} it is not possible to know that a failure will occur the instant before it actually occurs). Property \ref{monotonicity} asserts that information gained through discovery is not lost over time. Properties \ref{monotonicity} and \ref{convergence} acknowledge that one cannot be convinced that all useful modeling information will be revealed in finite time.\footnote{Even though it is reasonable to assert by definition that any consequential failure mode will be discovered in finite time, there is way to recognize that \emph{all} consequential failure modes will have been discovered by any historically observable time.}

Consider, now, the random variable $X_t: \Omega \rightarrow \{0, 1\}$ indicating the state of system protections where
	\begin{equation*}
		X_t =
   	 		\begin{cases}
        				1, & \text{protections are available at time }t\\
       				0, & \text{otherwise.}
    			\end{cases}
	\end{equation*}
We call $(X_t)_{t \ge 0}$ the \emph{protection availability process}, and we normatively take its trajectories to be right--continuous.\footnote{Because $(X_t)_{t \ge 0}$ is right--continuous with left--limits, it belongs to the class of c\`{a}dl\`{a}g ensuring that it has a left--continuous version with respect to the probability measure $P$.} We also take $(X_t)_{t \ge 0}$ to be adapted to the filtration $\{\mathscr{F}_t\}_{t \ge 0}$.  This is to say, at any time $t \ge 0$, $X_t$ is $\mathscr{F}_t$--measurable, and we have that, $\sigma((X_s)_{0 \le s \le t}) \subseteq \mathscr{F}_t$.\footnote{In other words, the ``natural history'' of the protection availability process is contained within the filtration capturing the flow of engineering information.} 
Of course, at any time $t \ge 0$, there is far more engineering information in $\mathscr{F}_t$ than simply the partial protection availability trajectories that comprise $\sigma(X_t)$. Information related to protection design, maintenance records, historical environmental conditions, \emph{etc} are also events that can appear in the filtration $\{\mathscr{F}_t\}_{t \ge 0}$.

For the purposes of predictive modeling, we must be careful to stipulate the status of engineering information. Note that the likelihood that system protections are failed at time $t$ is given by
\begin{equation}\label{total_probability}
	 P(X_t = 0) =  E[(1- X_t)] \triangleq E[(1- X_t) | \mathscr{F}]= E[E[(1 - X_t) | \mathscr{F}_t]].
\end{equation}
$\mathscr{F}_0$ is assumed to be complete and contains all information characterizing the protection design space with $t = 0$ taken as the time of deployment. So, all events associated with failure modes identified in an original design \ac{fmea} are included in $\mathscr{F}_0$.

\Cref{total_probability} reveals that quantifying the uncertainty about the operational integrity of protections at time $t$ requires information that is not 
necessarily available. To see this, examine the set difference $A_t \triangleq \mathscr{F} \setminus \mathscr{F}_t$. Intuitively, $A_t$ contains all as yet undiscovered engineering information at time $t$. It follows from \cref{total_probability} that $P(X_t = 0)$ can be quantified with respect to probability measure $P$ if and only if $P(A_t) = 0$. That is, we must be sure that there is no remaining undiscovered information that would be valuable in predicting the unavailability of protections.  $P(A_t) = 0$ is guaranteed only in the limit as stipulated in Property~\ref{convergence}, where a numerical value for $P(A_t) = 0$ is available only to a clairvoyant. Hence, engineering predictive modeling must be contented with quantifying uncertainty only up to the currently available engineering information, or
\begin{equation}\label{conditional_probability}
	 P(X_t = 0 | \mathscr{F}_t) = E[(1 - X_t) | \mathscr{F}_t].
\end{equation}

It is important to appreciate that system protections are useful only if they are available exactly when needed. Suppose that $T$ is the time of arrival of some initiating event that can possibly lead to catastrophe. Here, $T: (\Omega, \mathscr{F}) \rightarrow (\mathbb{R}^+, \mathcal{B}(\mathbb{R}^+))$.  Catastrophe can occur only if $X_T = 0$.
\begin{remark}
Simply characterizing the state of protections with respect to the flow of engineering information over time is generally inadequate to inform the likelihood of successful protections; protections must hold at the random times when initiating events occur. In practical circumstances, $P(X_T = 0) \neq P(X_t = 0)$, where $T$ is the arrival time of an initiating event.\footnote{This fact bears directly upon the validity of \ac{pq} methods when applied to risk informed, performance--based regulatory oversight.}
\end{remark}

	\subsection{Initiating event arrivals that are stopping times.}
When the events $\{T \le t\}$ are included in $\mathscr{F}_t$ for all $t \ge 0$, $T$ is said to be an $\{\mathscr{F}_t\}_{t \ge 0}$ stopping time.
It is well known that any $\{\mathscr{F}_t\}_{t \ge 0}$ stopping time $T$ must also be $\mathscr{F}_T$--measurable, where by definition
	\begin{equation}
		 \mathscr{F}_T \triangleq \{A \in \mathscr{F}: A \cap \{T \le t \} \in \mathscr{F}_t\}.
	\end{equation}
It follows directly that  $\mathscr{F}_T \subseteq \mathscr{F}_t$.  Thus, $P(X_T = 0 | \mathscr{F}_t) = E[(1-X_T) | \mathscr{F}_t]$ is well defined.

Clearly, stopping times play an important role in \ac{pq} supporting risk analysis and regulatory oversight. It should be appreciated that generally $\mathscr{F}_T$ contains engineering information that is \emph{not} necessarily included in the protection hardware design/maintenance space (\emph{e.g.,} information derived from prior understandings of weather patterns, political unrest, economic activity, \emph{etc}).
Under the circumstance that the arrival time of an initiating event is an $\mathscr{F}_t$--stopping time for all $t \ge 0$, we know that $P(X_T = 0 | \mathscr{F}_t)$ is well defined.  However, being well defined \emph{does not} imply that numerical values for $P(X_T = 0 | \mathscr{F}_t)$ are attainable \citep{wortman2021core}.

	\subsection{Initiating event arrivals that are \emph{not} stopping times.}
There are a host of practical reasons why the arrival time of an initiating event might not be measurable with respect to the filtration $\{\mathscr{F}_t\}_{t \ge 0}$.  In particular, if it happens that the initiating event reveals a previously undiscovered system protection failure mode, a catastrophe can ensue. Since the failure mode is heretofore unknown, its arrival time $T$ is \emph{not} $\mathscr{F}_t$--measurable for any $t \ge 0$. This is to say $E[(1 - X_T) | \mathscr{F}_t]$ is not well defined which implies that neither $P(X_T = 0 | \mathscr{F}_T)$ nor $P(X_T = 0 | \mathscr{F}_t)$ can be quantified. Thus, undiscovered protection failure modes are problematic for any regulatory oversight strategy that relies completely on \ac{pq}.

It is reasonable, at this juncture, to consider the extent to which protection failure modes not identified in $\mathscr{F}_0$ are problematic.  We begin with some observations that are normatively justified in the engineering pedagogy: 

\begin{itemize}
	\item The discovery time $T$ of a specific heretofore undiscovered failure mode is random and measurable with respect to probability space $(\Omega, \mathscr{F}, P)$.
	\item $T$ is not an $\{\mathscr{F}_t\}_{t \ge 0}$ stopping time.
	\item It is impossible to quantify the probability of events that depend on $T$.
	\item There exists some finite time $\tau < \infty$ such that $P(T \le \tau) = 1$.  That is, any consequential failure mode will almost surely be found over the lifecycle of protections, else cannot be consequential.
	\item It is possible that a failure mode is first discovered through catastrophe postmortem analysis (\emph{i.e.}, the failure mode caused catastrophe upon its first appearance).
	\item For any time $t \ge 0$ a non-clairvoyant cannot rule out the possibility of remaining undiscovered protection failure modes.
\end{itemize}
If undiscovered consequential failure modes are in play, the filtration $\{\mathscr{F}_t\}_{t \ge 0}$ cannot yet have converged. From the perspective of predictive modeling, it is understood that $A_t$ must contain the undiscovered failure mode information, and thus $P(A_t | \mathscr{F}) > 0$.  But, clearly, $A_t \notin \mathscr{F}_t$ and thus $P(A_t | \mathscr{F}_t)$ is not well defined ... that is, a non--clairvoyant is unable to quantify the inaccessible event probabilities related to undiscovered failure modes. 

However, if a modeler is sufficiently confident in a non-clairvoyant belief that that all consequential failure modes have been discovered, then all information impacting protection design will be found in the tail $\sigma$--algebra $\mathscr{T}$, where

	\begin{equation*}
		\mathscr{T} \triangleq \bigcap_{t \ge 0} \sigma((X_s)_{s > t}).
	\end{equation*}
By definition, $\mathscr{T} \subset \mathscr{F}$ and characterizes design information in the \emph{remote future} of $(X_t)_{t \ge 0}$. In the remote future of protections, there can be no undiscovered consequential failure modes.

Typically, \ac{pq} methodologies (\emph{e.g.,} \ac{pra}, \ac{qra}, \ac{psa}) implicitly rely on the assumption that events associated with protections are $\mathscr{T}$--measurable, because this assumption guarantees the existence of
	\begin{equation*}
		X = \lim_{t \rightarrow \infty} X_t \mbox{ and } \overline{X} = \lim_{t \rightarrow \infty} \frac{1}{t} \int_0^t X_sds.
	\end{equation*}
Expected values of these $X$ and $\overline{X}$ and their corresponding statistical estimators play essential roles in \ac{pq}.
Informally, when time $t=0$ is taken to be in the remote future of the protection availability process $(X_t)_{t \ge 0}$, then $\mathscr{T} \subset \mathscr{F}_t$ and $X$ and $\overline{X}$ each become $\mathscr{F}_t$--measurable, and numerical estimates of their respective expected values can (in principle) be computed using historical data collected up through time $t$.\footnote{Assuming that $(X_t)_{t \ge 0}$ is $\mathscr{T}$--measurable, in no way implies that an $\mathscr{F}_t$--measurable stopping time $T$ will be either $\mathscr{T}$--measurable or independent of $(X_t)_{t \ge 0}$.  For example an initiating event arriving at time $T$ could possibly damage system protections triggering a protection failure.  Thus, even when assuming that the protection availability process is $\mathscr{T}$--measurable, estimating $E[(1- X_T) | \mathscr{F}_t]$ can be perilous.}

Importantly, modelers should appreciate that enabling computation of \ac{pq} statistics nearly always requires assuming that $(X_t)_{t \ge 0}$ be $\mathscr{T}$--measurable.  And, assuming that $(X_t)_{t \ge 0}$ is $\mathscr{T}$--measurable implicitly ignores the possibility of undiscovered consequential failure modes that might lead to catastrophe.  This leaves open the engineering question, ``\emph{How much time must elapse before one might reasonably trust that all consequential protection failure modes have been discovered?}''. Mathematics dictates that only a clairvoyant can answer this question with complete confidence.
Of course, no one is clairvoyant and undiscovered protection failure modes present especially difficult problems for engineers. Hence, because protection failure mode discovery times defy \ac{pq}, care must be exercised when applying \ac{pq} methodologies to studies of protection efficacy.

\section{Discussion}

Unknown--unknowns, including undiscovered protection failure modes, are nothing new to protective system design.  Design practices such as stress testing, accelerated life testing, burn--in, and advanced physics--based simulations have been developed and refined with the specific objective of discovering failure mechanism.  Recognizing the existence of potentially catastrophic operating conditions that initially escape \ac{pq}, protection designers routinely adopt deterministic strategies that include defense in depth and layers of protection for mitigating the consequences of unknown--unknowns.  Additionally, the \ac{nrc} has deployed data recording and sharing protocols specifically intended to capture and newly discovered failure mechanisms so that these can be quickly shared across the entire nuclear industry. 
	
\ac{pq} based methodologies have been increasing in popularity in many areas of cost--benefit decision making in particular, regulatory oversight. The language appearing in \ac{neima} is an example of federal legislation promoting a preference for a risk informed, performance--based  approach to licensing of (new) advanced reactors, specifically appealing to the use of \ac{pra} - an example of \ac{pq} methodology that ignores the unknown--unknown failure modes that may exist  in advanced reactor protections where operating experience is limited. Development of the \ac{neima} legislation is supported by \ac{nei}, the industry trade and lobbying organization and reveals how well--intended political objectives can potentially raise the potential for severe economic loss through misapplied analytics.
Even in circumstances where there is a long history of operating experience, \ac{pq} methods such as \ac{pra} produce optimistic estimates of predicted protection performance that cannot be overcome \citep{wortman2021core}.

The validity of any \ac{pq}--based methodology is guaranteed only insofar as its application does not conflict with probability theory.  We emphasize that \ac{pq} applied to characterizing the efficacy of safety--critical protections is especially vulnerable to confounds in validity.  Hence, we are especially concerned by the  \ac{neima} legislative language (quoted in \Cref{introduction}, \Cpageref{neima_language}) which suggests that a risk informed, performance--based regulatory approach is capable of delivering insights that are actually beyond \ac{pq} validity.
Finally, although the example we show is from the nuclear power industry, our observations apply similarly to other hazardous industrial sectors such as chemical processing, oil and gas exploration and production, public and freight transportation, and others.
	%

\bibliography{bibtex_library_dec9}

\end{document}